\documentclass[10pt]{article}

\usepackage{epsfig}
\usepackage{amssymb}
\usepackage{amsmath}
\usepackage{amsthm}
\usepackage{verbatim}
\usepackage{multirow}
\usepackage{array}

\renewcommand{\baselinestretch}{2.0}    
\begin{document}

\pagestyle{plain}
\pagenumbering{arabic}
\setcounter{page}{1}

\begin{flushright}
\end{flushright}

\begin{center}

{\Large \bf ODYS: A Massively-Parallel Search Engine
} \\
\vspace*{-0.05cm} {\Large \bf Using a DB-IR Tightly-Integrated
Parallel DBMS}

Kyu-Young Whang$^{1}$, Tae-Seob Yun$^{1}$, Yeon-Mi Yeo$^{1}$, Il-Yeol Song$^{2}$, \\ Hyuk-Yoon Kwon$^{1}$, and In-Joong Kim$^{1}$ \\
\vspace*{-0.20cm}
$^{1}$ Department of Computer Science, KAIST, Daejeon, Korea, \\ \{kywhang, tsyun, ymyeo, hykwon, ijkim\}@mozart.kaist.ac.kr \\
\vspace*{-0.20cm}
 $^{2}$ College of Information Science and Technology, Drexel University, Philadelphia, USA, \\
\vspace*{-0.20cm}
songiy@drexel.edu
\end{center}


\renewcommand{\baselinestretch}{1.6}
\begin{abstract}
{\small Recently, parallel search engines have been implemented
based on scalable distributed file systems such as Google File
System. However, we claim that building a massively-parallel
search engine using a parallel DBMS can be an attractive
alternative since it supports a higher-level (i.e., SQL-level)
interface than that of a distributed file system for easy and less
error-prone application development while providing scalability.
In this paper, we propose a new approach of building
a massively-parallel search engine using a DB-IR
tightly-integrated parallel DBMS and demonstrate its
commercial-level scalability and performance. In addition, we
present a hybrid (i.e., analytic and experimental) performance model for the
parallel search engine.
We have built a five-node parallel search engine according to the proposed
architecture using a DB-IR tightly-integrated DBMS\,\cite{Wha05}.
Through extensive experiments, we show the correctness of the
model by comparing the projected output with the experimental
results of the five-node engine. Our model demonstrates that ODYS
is capable of handling 1 billion queries per day (81 queries/sec) for 30 billion
web pages by using only 43,472 nodes with an average query response
time of 211 ms, which is equivalent to or better than those of
commercial search engines. We also show that, by using twice as
many (86,944) nodes, ODYS can provide an average query response time of 162 ms,
which is significantly lower than those of commercial search engines.
}
\end{abstract}
\renewcommand{\baselinestretch}{2.0}

\newtheorem{lemma}{\bf Lemma}
\newtheorem{theorem}{\bf Theorem}
\newtheorem{algorithm}{\sc Algorithm}

\theoremstyle{definition}
\newtheorem{definition}{Definition}
\newtheorem{example}{Example}

\section{Introduction}

%
%
\subsection{Motivation}
\label{subsec:Motivation}

A Web search engine is a representative large-scale system, which
handles billions of queries per day for a petabyte-scale database
of tens of billions of Web pages\,\cite{Dea09,Gul05,Kun}. Until
now, commercial Web search engines have been implemented based on
a scalable distributed file system such as Google File System
(GFS)\,\cite{Ghe03} or Hadoop Distributed File System
(HDFS)\,\cite{HDFS}. These distributed file systems are suitable
for large-scale data because they provide high scalability using a
large number of commodity PCs. A storage system proposed for
real-world-scale data with better functionality is the key-value
store. It stores data in the form of a key-value map, and thus, is
appropriate for storing a large amount of sparse and structured
data. Representative key-value stores are Bigtable\,\cite{Cha06},
HBase\,\cite{HBa}, Cassandra\,\cite{Cas}, Azure\,\cite{Azu}, and
Dynamo\,\cite{DeC07}. These systems are based on a large-scale
distributed storage such as a distributed file
system\,\cite{Cha06,HBa} or a distributed hash
table(DHT)\,\cite{Azu,Cas,DeC07}.

However, both distributed file systems and key-value stores, the
so-called ``NoSQL'' systems, have very simple and primitive
functionalities because they are low-level storage systems. In
other words, they do not provide database functionalities such as
SQL, schemas, indexes, or query optimization. Therefore, to
implement high-level functionalities, developers need to build
them using low-level primitive functions. Research for developing
a framework for efficient parallel processing of large-scale data
in large storage systems has been proposed.
MapReduce\,\cite{Dea04} and Hadoop\,\cite{Had}
are the examples of parallel processing
frameworks. These frameworks are known to be suitable for
performing extract-transform-load (ETL) tasks or complex data
analysis. However, they are not suitable for query processing on
large-scale data because they are designed for batch processing
and scanning of the whole data\,\cite{Sto10}. Thus, commercial search
engines use these frameworks primarily for data loading or
indexing instead of user query processing.

\begin{figure}[hbt]
\centerline{\epsfig{file=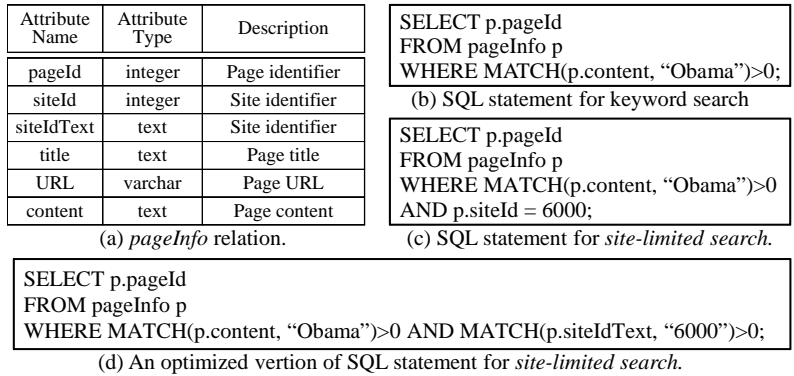,width=10.5cm}}
\vspace*{-0.5cm} \caption{An example of a schema and SQL
statements.} \label{fig:schema and SQL}
\end{figure}

The high-level functionalities such as SQL, schemas, or indexes
that are provided by DBMSs allow developers to implement queries
that are used in search engines easily because they provide a
higher expressive power than primitive functions in key-value
stores, facilitating easy (and much less error-prone) application
development and maintenance. In this sense, there have been many
research efforts to support SQL even in the NoSQL
systems\,\cite{Cha08,Thu09}. Fig.~\ref{fig:schema and SQL} shows the representative queries
used in search engines that are simply specified using the
high-level functionalities. Fig.~\ref{fig:schema and SQL}\,(a)
shows a schema of a relation \textit{pageInfo} that represents the
information of Web pages. Fig.~\ref{fig:schema and SQL}\,(b)
shows a SQL statement that represents a keyword query. The query
finds the Web pages that contain the word ``Obama'' from the
\textit{pageInfo} relation. Fig.~\ref{fig:schema and SQL}\,(c)
shows a SQL statement that represents a site-limited search;
Fig.~\ref{fig:schema and SQL}\,(d) shows one of its optimized
versions to be explained in Section~\ref{subsec:DB-IR_Integration}. \textit{Site-limited search} limits the scope of a user
query to the set of Web pages collected from a specific
site\,\cite{Wha05}. The query finds the Web pages that contain the
word ¡°Obama¡± from the site having siteId 6000.

\begin{figure}[hbt]
\centerline{\epsfig{file=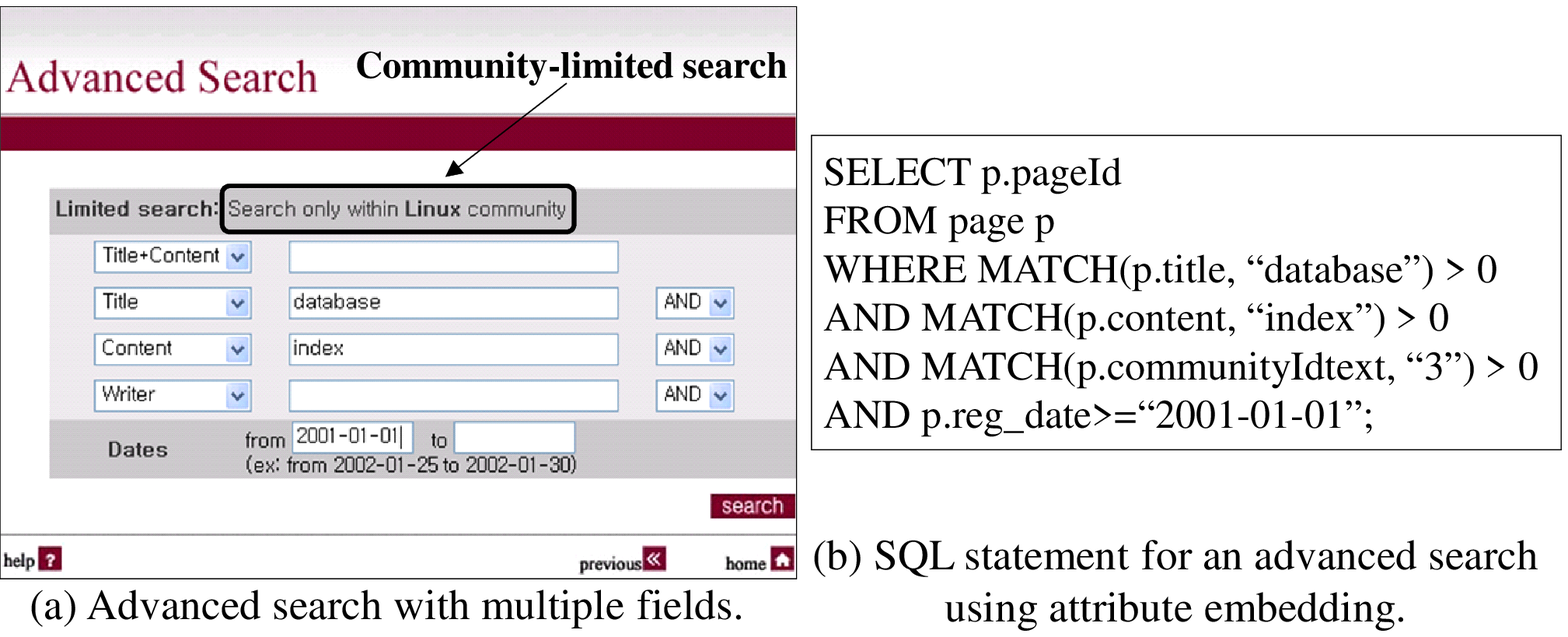,width=13.5cm}}
\vspace*{-0.5cm} \caption{An example of advanced search.}
\label{fig:advanced_search}
\end{figure}

These high-level functionalities allow us to easily develop
advanced search engines with multiple search fields such as
on-line discussion board systems as shown in
Fig.~\ref{fig:advanced_search}\,(a). The presented advanced
search involves multiple fields as well as community-limited search capability.
It requires complex join operations among multiple
fields, which require implementations with high-level complexity.
However, SQL allows us to implement those operations with simple specification.
Fig.~\ref{fig:advanced_search}\,(b) shows a simple SQL statement for an
advanced search. Likewise, other search related applications can be easily developed by using SQL.

A parallel DBMS is a database system that provides both storage
and parallel query processing capabilities. It could be considered an
alternative to a large-scale search engine because it has higher
scalability and performance than traditional single node DBMSs and
also has rich functionality such as SQL, schemas, indexes, or
query optimization. Stonebreaker et al.\,\cite{Sto10} argue that
parallel DBMSs are scalable enough to handle large-scale data and
query loads. They claim that parallel DBMSs are linearly scalable
and can easily service multiple users for database systems with
multi-petabytes of data. However, parallel DBMSs have been
considered as not having enough performance and scalability to be
used as a large-scale search engine\,\cite{Abo09,Dea04}, one
outstanding reason being lack of efficient information retrieval
(IR) functionalities.

To enable a DBMS to efficiently handle keyword search, Whang et
al.\,\cite{Wha02,Wha05} have proposed tight integration of
database (DB) and information retrieval (IR) functionalities.
Tight DB-IR integration implements IR functionalities within the
core of a DBMS, and thus, IR performance becomes efficient due to
short access paths to the IR functionalities. Whang et
al.\,\cite{Wha02,Wha05} also present two efficient techniques for
providing DB-IR integrated queries: 1) IR index join with posting
skipping\,\cite{Guo03,Hal03,Wha02,Wha05} and 2) attribute
embedding\,\cite{Wha05} to be explained in Section~\ref{subsec:DB-IR_Integration}.

%
%

\subsection{Our Contribution}
\label{subsec:Contribution}

In this paper, we make the following three
contributions. First, we show that we can construct a
commercial-level massively parallel search engine using a parallel
DBMS, which to date has been considered infeasible.
The proposed architecture ODYS, featuring a shared-nothing
parallel DBMS, consists of masters and slaves.
ODYS achieves commercial-level scalability and efficiency by using
Odysseus that has DB-IR tight integration functionalities\,\cite{Wha05} as its slaves.
We have verified that each Odysseus is capable of indexing 100
million Web pages (loading and indexing in 9.5 days in a LINUX
machine\footnote{The machine is with a quad-core 2.5 GHz CPU, 4
Gbytes of main memory, and a RAID 5 disk having 13 disks (disk
transfer rate: average 83.3 Mbytes/s) with a total of 13 Tbytes, a
cache of 512 Mbytes, and 512 Mbytes/s bandwidth.}), and thus, our
parallel DBMS is capable of supporting a large volume of data with
a relatively small number of machines\footnote{Typical commercial
search engines index 20 million Web pages per node (or shard), and
thus, require five times as many machines.}. Furthermore, tight
integration of DB-IR functionalities enables the architecture to
efficiently process a large number of queries arriving at a very
fast rate. We show that our parallel DBMS can achieve a
commercial-level performance especially for single-keyword
searching, which is the most representative query.

Second, we present a sophisticated analytic and
experimental performance model (simply, a \textit{hybrid model})
that projects
the performance of the proposed architecture of the parallel DBMS,
and then, validate the accuracy of the model. For the master and
network, we model each system component using the queuing model,
and then, estimate the performance.
For the slave, we propose an experimental method for
accurately predicting the performance of a scaled-up system (e.g.,
300-node) using a small-scale one (e.g., 5-node). We note that the
bulk (say, 85.36\%\,$\sim$\,93.47\%) of the query processing time is spent at the
slave compared with the master and network. Our experimental
method ensures high predictability of the slave side query
processing time since the estimation is directly derived from the
actual measurement.
To verify the correctness of the model,
we have built a five-node parallel system of the proposed architecture
and perform experiments with query loads compatible to those of a
commercial search engine. The experimental results show that the
estimation error between the model and the actual
experiment is less than 0.59\%. The proposed model
allows us to substantially reduce costs and efforts in building a
large-scale system because we can estimate its performance using a
small number of machines without actually building it.

Last, by using the performance model, we
demonstrate that the proposed architecture is capable of handling
commercial-level data and query loads with a relatively
small number of machines.
Our result shows that, with only 43,472 nodes, the proposed
architecture can handle 1 billion
queries/day\footnote{Nielsenwire\,\cite{Nie10} reports that Google
 handled 214 million queries/day in the U.S. in February 2010.} for 30 billion Web pages with an average query response time of
211 ms while commercial search engines typically use hundreds of
thousands nodes\,\cite{Dea09b} for this query load and data
volume. We also show that, by using twice as many (i.e.,
86,944) nodes, ODYS can provide an average query response time of
162 ms, which is significantly lower than those of commercial
search engines\footnote{The announced average query latency of
Google is 0.25 seconds~\cite{Goo}.}. This clearly demonstrates the scalability and efficiency of our
architecture.

The rest of this paper is organized as follows.
Section~\ref{sec:Preliminary}
introduces techniques of DB-IR integration as a preliminary.
Section~\ref{sec:ODYS} proposes the architecture of ODYS, a
massively parallel search engine using a DB-IR tightly integrated
parallel DBMS. Section~\ref{sec:Performance_Estimation_Model}
proposes the performance model of ODYS.
Section~\ref{sec:Performance_Evaluation} presents the experimental
results that validates the proposed performance model and
demonstrates the scalability and performance of ODYS.
Section~\ref{sec:Conclusions} concludes the paper.

\vspace*{-0.5cm}
\section{DB-IR Integration}
\label{sec:Preliminary}

%
%
\label{subsec:DB-IR_Integration}
In the database (DB) research field, integration
of DBMS with information retrieval (IR) features (simply,
\textit{DB-IR integration}) has been studied actively as the need
of handling unstructured data as well as structured data is
rapidly increasing\,\cite{Bae04,Cha05,Wha02,Wha05}.
There are two approaches to DB-IR integration, loose coupling and
tight coupling. The loose coupling method---used in the commercial
systems---provides IR features as user defined types and functions
outside of the DBMS engine (e.g., Oracle Cartridge and IBM
Extender). This method is easy to implement because there is no
need to modify the DBMS engine, but the performance of the system
gets degraded because the access path for the IR feature becomes
long. On the other hand, the tight coupling method proposed by
Whang el al.\,\cite{Wha02,Wha05,Wha11} directly implements data
types and operations for IR features as built-in types and
functions of a DBMS engine
(e.g.,Odysseus\,\cite{Wha02,Wha05,Wha11} and MySQL\,\cite{Len04}).
The implementation of the method is difficult and complex because
the DBMS engine should be modified, but the performance increases.
Thus, the tight coupling method is appropriate for a large-scale
system to efficiently handle a large amount of data and high query
loads. The IR index of Odysseus\,\cite{Wha02}\footnote{Patented in
the US in 2002; application filed in 1999.} and MySQL are very close, but Odysseus has more
sophisticated DB-IR algorithms for IR features as discussed below.

\begin{figure}[hbt]
\centerline{\epsfig{file=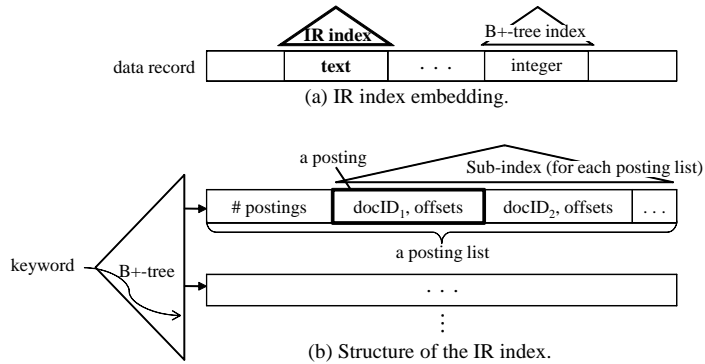, width=9.5cm}}
\caption{The IR index of the DB-IR tightly integrated DBMS.}
\label{fig:IR_index_of_DB_IR_tight}
\end{figure}

In a tightly integrated DB-IR system, an IR index
is embedded into the system as shown in
Fig.~\ref{fig:IR_index_of_DB_IR_tight}\,(a). As in a typical
DBMS, a B+-tree index can be constructed for an integer or
formatted column. Similarly, an IR index is (automatically)
constructed for a column having the text type.
Fig.~\ref{fig:IR_index_of_DB_IR_tight}\,(b) shows the structure
of the IR index. The IR index consists of a B+-tree index for the
keywords, where each keyword points to a \textit{posting list}.
The leaf node of the B+-tree has a structure similar to that of an
inverted index. Each posting list for a keyword consists of the
number of postings and the postings for the keyword. A
\textit{posting} has the document identifier (docID) and the
location information where the keyword appears (i.e., docID,
offsets). On the other hand, different from the inverted index,
the IR index has a sub-index\,\cite{Wha02}$^5$ for each posting list
to search for a certain posting efficiently.

In the DB-IR tightly integrated DBMS, two methods are used to
improve the search performance: IR index join with posting
skipping\,\cite{Guo03,Hal03,Wha02,Wha05} (also called the ZigZag join\,\cite{Hal03})
and attribute embedding\,\cite{Wha05}. IR index join with posting skipping is a
technique for efficiently searching documents that have multiple
co-occurring keywords. To search for documents having co-occurring
keywords, the posting lists of the keywords should be joined. The
posting skipping method identifies the part of the posting lists
that need to be merged and skips the rest by using sub-indexes.
Attribute embedding is a technique for efficiently processing a
DB-IR query that joins an attribute of a structured data type and
an attribute of the text data type. For example, suppose that there
are two attributes A and B having the text type and the integer
type, respectively, and they are often accessed together. The
attribute embedding method embeds the value of attribute B in each
posting of attribute A. In this case, a DB-IR query that joins
attributes A and B can be simply processed by one sequential scan
of the posting list. In summary, it is the tightly integrated IR
features, such as the embedded IR index with posting skipping, and
attributed embedding, that makes ODYS a powerful search engine in
the proposed parallel DBMS.

\begin{figure}[hbt]
\centerline{\epsfig{file=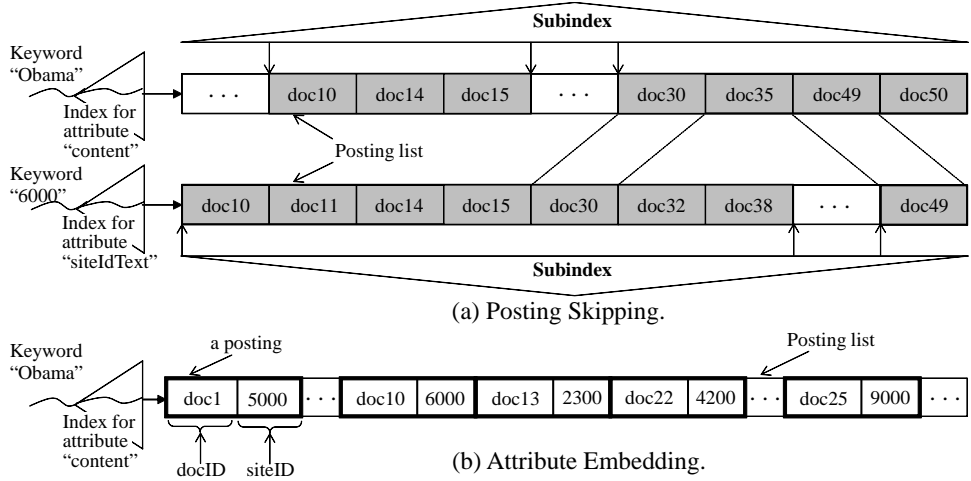,
width=13cm}} \caption{IR query processing in a tightly integrated
DB-IR system~\cite{Wha05}.}
\vspace*{-0.7cm}
\label{fig:IR_index_join_attributed_embedding}
\end{figure}

\begin{example}
Fig.~\ref{fig:IR_index_join_attributed_embedding} shows the
processing of an IR query in a tightly integrated DB-IR system.
Fig.~\ref{fig:IR_index_join_attributed_embedding}\,(a) shows an
example of IR index join with posting skipping. When a
site-limited query as in Fig. 1\,(c) is given, $siteIdText$ of
type text is used instead of $siteId$ of type integer as in Fig.~1\,(d).
Thus, the postings to be merged for each posting list are
found efficiently using sub-indexes, as in the multiple
keyword query processing. Fig.~\ref{fig:IR_index_join_attributed_embedding}\,(b) shows an
example of attribute embedding. The values for
$siteId$ of type integer are embedded in the postings of
$Content$. We can efficiently process the queries involving both
$siteId$ and $Content$ as in Fig. 1\,(c) by one sequential scan
of the posting list. Attribute embedding is easily specified by schema declaration using an SQL-like language\,\cite{Wha11}.$\Box$
\end{example}

\section{ODYS Massively-Parallel Search Engine}
\label{sec:ODYS}

%
%
\subsection{Architecture}
\label{subsec:Architecture}

\begin{figure}[hbt]
\centerline{\epsfig{file=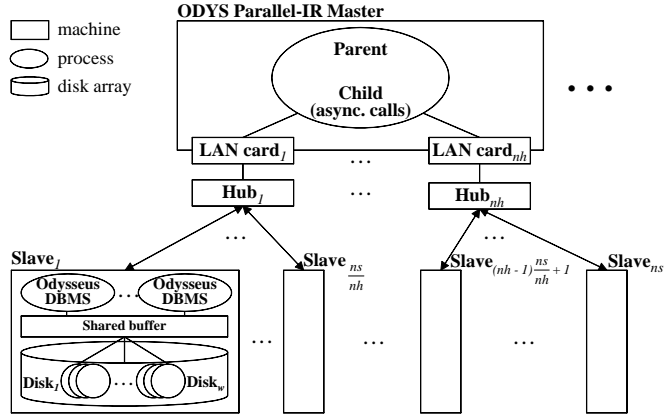, width=9cm}}
\caption{The architecture of ODYS.}
\label{fig:Architecture}
\end{figure}

ODYS is a massively-parallel DBMS capable of managing tens of
billions of Web pages. Fig.~\ref{fig:Architecture} shows the
architecture of ODYS. ODYS consists of masters\,\footnote{The ODYS
Parallel-IR Master consists of 58,000 lines of C and C++ code.}
and slaves\,\footnote{The Odysseus DBMS (slave) consists of
450,000 lines of C and C++ code.}. The masters share the slaves,
and the slaves have a shared-nothing architecture. Each slave is Odysseus storing data in a disk array.
The master and the slaves are connected by a gigabit network hub,
and they communicate by using an asynchronous remote procedure
call (RPC)\footnote{We use socket-based RPC consisting of 17,000
lines of C, C++, and Python code developed by the authors.}.

The master stores metadata such as slaves' IP addresses, slaves'
database paths, and schemas. The slaves store crawled Web pages
and their IR indexes. There are two well-known methods for
partitioning the index~\cite{Dea09}: 1) \textit{partitioning by
documents} and 2) \textit{partitioning by keywords}.
For performance reasons, most commercial search engines including
Google use the former method~\cite{Dea09}, which makes slaves work
in parallel for processing the same query. Thus, we also employ
the same method. That is, the entire set of Web pages is
partitioned horizontally. Each slave stores a segment of the
partitioned data and creates an IR index for each text-type
attribute in the segment.

ODYS processes a user query as follows. When a user query arrives
at a master, the master distributes the query to all slaves. Then, the master merges the
results returned from slaves and returns the final results to the
user. The slaves process the query and return the results to the
master. Each slave returns top-$k$ results in the ranking order,
and the master performs a merge operation for the results returned
from the slaves to get the final top-$k$ results. The slaves store
each posting list of the IR index in the PageRank~\cite{Bri04}
order (i.e., we are using query-independent ranking~\cite{Lan06})
to make the top-$k$ search efficient. Since the posting lists are
stored in the PageRank order, the query processing performance is not
much affected regardless how long the posting lists are or how big
the database is. In this paper, we focus on the performance issues
of the search engines and not on the effectiveness of ranking results.
For performance reasons, we use query-independent ranking,
which can be computed in advance, as many large-scale
commercial search engines do, but any other ranking methods can be
applied.

In the current version of ODYS, fault tolerance functionalities
have not yet been implemented. However, this could be accomplished
by adopting the approach proposed in Osprey\,\cite{Yan10}.
In Osprey, Yang et al.\,\cite{Yan10} proposed a method
implementing MapReduce-style fault tolerance functionalities to
the parallel DBMS. The proposed method maintains replicas and
allocates a small sized tasks dynamically to the nodes according
to the loads of each node. As in Osprey, the availability and
reliability can be improved by maintaining multiple replicas of
ODYS, and a middleware can be used for dynamically mapping masters
and slaves of the multiple ODYS replicas. We call a replica of
ODYS an \textit{ODYS set}.

Updates among multiple nodes in distributed systems require
complex consistency protocols such as two-phase commit to
guarantee strong consistency; consequently, they incur
degradation of system performance. Thus, commercial search
engines do not support immediate updates involving multiple nodes;
instead, they replace existing nodes periodically with new
nodes based on updates by each node~\cite{Bar03}. ODYS supports
updates on a per-node basis with strong consistency~\cite{Wha11}
so any transaction pertaining to individual nodes can be properly handled~\cite{Wha05}.
In this paper, we focus on the retrieval performance issues of search engines,
and the detailed discussion on updates is omitted.

%
%
\subsection{Other Parallel Processing Systems}
\label{subsec:Other_Parallel_Processing_Systems}

\begin{figure*}
\vspace*{0.50cm}
\centering \epsfig{file=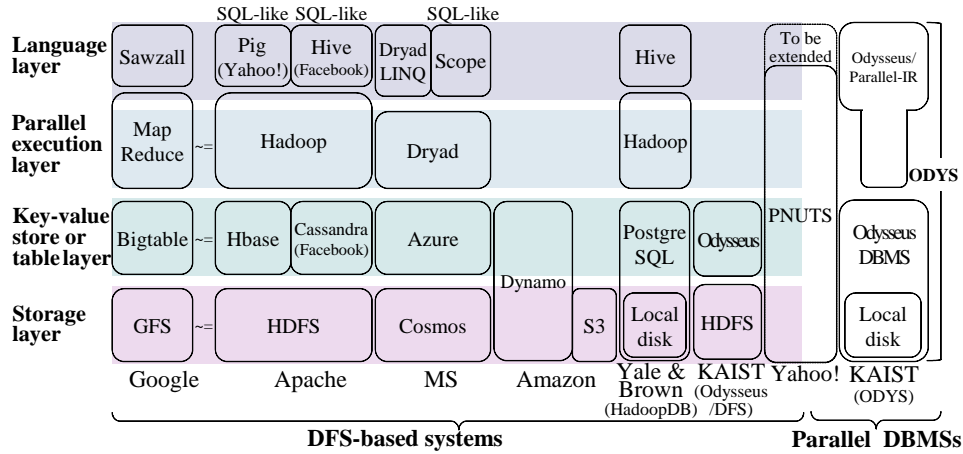, width=13cm} \caption{The map of ODYS and other parallel
processing systems.} \label{fig:Map_of_ODYS_and_other_PPS}
\end{figure*}

In this section, we discuss the architectural relationships between
ODYS and other recently developed parallel processing systems. We
classify the existing DFS-based systems and parallel DBMSs into
four types of layers as shown in
Fig.~\ref{fig:Map_of_ODYS_and_other_PPS}\,\footnote{Modified and
extended from the figure in p.4 of \cite{Bud09}.}. In
Fig.~\ref{fig:Map_of_ODYS_and_other_PPS}, the storage layer
represents a distributed storage for large-scale data. The
key-value store or a table layer represents a data storage storing
data in the form of key-value pairs or records in tables. The
parallel execution layer represents a system for automatically
parallelizing the given job. The language layer represents a
high-level query interface.

Most DFS-based systems that have been recently developed follow or
modify the Google's architecture. DFS-based systems developed by
Apache and Microsoft (MS) have an architecture very close to
Google's. On the other hand, Amazon's Dynamo, HadoopDB, and
Odysseus/DFS can be considered as variations of Google's
architecture. Dynamo has a fully decentralized architecture and
uses a relaxed consistency model called eventual consistency.
HadoopDB is a Hadoop on top of multiple single-node DBMSs while
Odysseus/DFS is a relational DBMS on top of HDFS.
PNUTS is a highly scalable parallel DBMS that provides carefully
chosen functionalities. It shares some design choices with the
DFS-based system in that it provides simple functionalities, a
relaxed consistency model, and flexible schema.

ODYS consists of two layers: Odysseus DBMS and Odysseus/Parallel-IR.
The Odysseus DBMS corresponds to the table layer. In ODYS, Odysseus
DBMSs with local disks are used in parallel rather
than key-value stores with a DFS system. The Odysseus/Parallel-IR
is an integrated layer that combines the parallel execution layer
and the language layer. Because ODYS uses a DBMS for the table
layer, it can provide rich functionality for query processing by
directly using most functionalities of the DBMS including SQL.

There are several open source projects for search engines. Solr
and Nutch are parallel search engines based on Apache
Lucene~\cite{Luc}, which is a search engine library for a single
machine. They have a similar architecture consisting of masters
and slaves where each slave is built on HDFS or a local file system.
However, they do not support the high-level language layer
such as SQL, but only support keyword queries. In addition, their
performance and scalability are not enough to support
commercial-level query loads. Moreira et al.~\cite{Mor07} analyze
performance and scalability of Nutch. The experimental results
indicate that Nutch can process 33 queries/second (i.e., 2.8
million queries/day) for 480 GBytes dataset using 12
machines~\cite{Mor07}\footnote{Moreira et al.~\cite{Mor07}
do not specify the ranking measure used. However, it must be that
they used query-dependent ranking (e.g., TF-IDF) since their results are
significantly (8\,$\sim$\,9 times) worse than those of 5-node ODYS to be
explained in Section~\ref{subsubsec:Accuracy_of_the_Performance_Model}. Thus, we do
not directly compare their results with ours here.}.

\section{Performance Model}
\label{sec:Performance_Estimation_Model}

In this section, we present a performance model of the proposed
search engine architecture. Except for the specialized search
engine companies, it is very difficult for a research center or an
academic institute to build a real-world-scale search engine
because of limited resources including availability of hardware,
space, and manpower. Therefore, an elaborate analytic or
experimental model is needed to test and project the performance
and the scalability of a large-scale system without actually
building it. For massively-parallel processing systems, analytic
models using the queuing theory have been proposed to estimate the
performance of the systems\,\cite{Jav04,Sha00}. However, those
analytic models cannot be simply applied to our architecture
because of the following three reasons. First, the existing
methods use simple parameters. In practice, however, to accurately
estimate the performance of a large-scale system, all the specific
parameters related to the performance of the system should be
identified. Second, the existing methods assume that there is only
one query type, while we consider multiple types of queries. Last,
as the phenomenon that most significantly affects the performance,
we show that the query response time is bounded by the maximum
among slaves' query processing times, but no existing analytic method takes
this into account. Therefore, we propose a performance model based
on the queuing theory as well as intricate measurement schemes.

We claim that our performance model using a small-scale system
(i.e., 5-node) can quite accurately predict the performance of a
large-scale system (i.e., 300-node) due to the following reasons.
The performance model consists of two parts: 1) the master and
network time and 2) the slave time. We show that the estimation
error of the former is very low, i.e., maximum 3.62\% as shown in
Fig.~\ref{fig:Exp_real_world_simulation} in
Section~\ref{subsubsec:Accuracy_of_the_Performance_Model}.
Moreover, even if the estimation error of the master and network
time were sizable, it could not affect the overall performance in
a significant way since the overall performance largely depends on
the performance of the slave time (e.g., 93.47\% for 15.5 million
queries/day) as shown in
Fig.~\ref{fig:Exp_real_world_scale_estimation}. We can be assured that
the estimated performance of slaves is very close to the actual
measurement since the estimation is directly derived from the
measurement as presented in Section
~\ref{subsec:The_Estimation_Method_of_the_Expected_Slave_Max_Time}.
Thus, our estimation of the total query response time is quite
accurate (e.g., the estimation error is less than 0.59\%) as shown
in Fig.~\ref{fig:Exp_real_world_simulation}.

The assumptions related to query execution are as follows.
We assume that every query is a top-$k$ query where $k$ is one of 10,
50 or 1000, and the set of input queries are a mix of
single-keyword queries, multiple-keyword queries, and limited
search queries as will be explained in Section~\ref{subsubsec:Query_Model}.
We also assume that every query is performed at semi-cold start.
\textit{Semi-cold start} means that a query is executed in the circumstance
where the internal nodes of the IR indexes (which normally fit in main
memory) are resident in main memory while the leaf nodes (which
normally are larger than available main memory), posting lists,
and the data (i.e., crawled Web pages) are resident in disk.
Typical commercial search engines process queries at \textit{warm start}
by storing all the indexes in a massive-scale main memory
\,\cite{Dea09}. This significantly helps reduce query response time.
However, to evaluate a lower-bound performance of our system,
we take a very conservative assumption: we run ODYS at semi-cold start.
To enforce semi-cold start, we use a buffer of only 12 MBytes sufficient
for containing the internal nodes (occupying 11.5 MBytes) of the IR index for each slave\footnote{
In our system, the size of the IR index is 2.81 Gbytes,
out of which leaf nodes occupy 2.8 Gbytes and the
internal nodes 11.5 Mbytes for indexing 114 million Web pages.}.

In the proposed performance model, each system component is
modeled as a queue, and the query response time is estimated by
summing up the expected sojourn time \,\cite{Coo81} of each queue.
The following are the major considerations of the performance
model. First, since each system component executes various types
of tasks, the queue representing the component would receive
various types of tasks that take different processing times. For
example, a master CPU performs tasks of distributing the query to
the slaves, merging the results returned from the slaves, etc. To
simplify the problem, however, we regard the summation of all the
types of tasks for a component as one request for the
corresponding queue.
Second, the service time is different according to search
condition types and the value of $k$ of the top-$k$ query. (The
service time is the time for a request to be serviced in the queue
excluding the waiting time.) For instance, the time taken by a
master CPU increases as the value of $k$ gets larger because the
merging cost increases. Therefore, we adopt the single-keyword
top-10 query type, which has the shortest processing time, as the
unit query and transform other query types in terms of the unit
query. We explain this transformation in detail in
Section~\ref{subsec:Queuing_Model}.
Finally, the times taken by slaves are bounded by the maximum
value among all the sojourn times of slaves. In other words, we
should calculate the expected maximum value of multiple sojourn
times. However, this estimation is known to be very hard in the queuing
theory\,\cite{Kem09}. Thus, we propose an estimation method for
the slave maximum time through experiments, and it will be
discussed in detail in
Section~\ref{subsec:The_Estimation_Method_of_the_Expected_Slave_Max_Time}.
\vspace*{-0.40cm}

%
%
\subsection{Queuing Model}
\label{subsec:Queuing_Model}

\vspace*{-0.30cm}
Among the ODYS system components, times taken by master CPUs,
master memory bus, and network hubs are estimated by using a
queuing model. Table~\ref{tbl:Symbol_table} shows the notation used in the performance model.

\renewcommand{\baselinestretch}{1.10}
\begin{table}[b!]
\begin{center}
\caption{The notation used in the performance model.}
\label{tbl:Symbol_table} \vspace*{0.1cm}
{\small  
\begin{tabular}[!b]{|c|m{6.5cm}|} \hline
\makebox[1.5cm]{{Symbols}} & \multicolumn{1}{c|}{{Definitions}} \\
\hline \hline
$nm$ & the number of master nodes  \\
\hline
$ncm$ & the number of CPUs per master  \\
\hline
$ns$ & the number of slave nodes  \\
\hline
$nh$ & the number of network hubs  \\
\hline
$\lambda$ & the arrival rate of the ODYS \\
\hline
$\lambda_{C}$ & the arrival rate of the system component $C$   \\
\hline
$\lambda_{C}'$ & the weighted arrival rate of the system component $C$ \\
\hline
$w_{C}(k)$ & the weight of the top-$k$ query type in the system component $C$   \\
\hline $qmr(sct,k)$ & the query mix
ratio of top-$k$ queries where $sct$ is the search condition type \\
\hline
$ST_{C}$ & the service time of the system component $C$   \\
\hline
$L_{C}$ & the average queue length of the system component $C$   \\
\hline
$X_{C}$ & the sojourn time of a customer in the queue of the system component $C$   \\
\hline
$r$ & the number of repetitions of a query set execution \\
\hline
$np$ & the number of slaves in the small-sized system built \\
\hline
$m_{i}$ & the master processing time for the $i$th slave \\
\hline
$s_{i}$ & the query processing time of the $i$th slave  \\
\hline
$nt_{i}$ & the network transfer time of the $i$th slave's results \\
\hline
\end{tabular}
} 
\vspace*{-0.5cm}
\end{center}
\end{table}
\renewcommand{\baselinestretch}{2.0}

%
%
\subsubsection{Query Model}
\label{subsubsec:Query_Model}

We consider nine types of queries.
First, the queries are classified into three search condition
types: single-keyword, multiple-keyword, and limited search queries.
For each of the three search condition
types, we further consider three types of top-$k$
queries, where $k$ = 10, 50, and 1,000. The performance of the
master CPUs, master memory bus, and network hubs depend only on
the $k$ value while the performance of the components in slaves
depends on both the search condition type and the $k$ value.

A \textit{single-keyword query} is a query that
has one keyword search condition. It is processed by finding the
posting list that corresponds to the keyword from the IR index and
by returning the first $k$ results of the posting list. Single-keyword
queries having the same $k$ value can be processed within almost
the same time regardless of the keyword because
the top-$k$ results have been stored according to a query-independent ranking
calculated apriori.

A \textit{multiple-keyword query} is a query that has two or more
keyword search conditions. It is processed by finding the posting
list for each keyword, performing a multi-way join for the posting
lists, and returning the first $k$ results of the joined results.
A \textit{limited search query} is a query having a keyword
condition together with a site ID or a domain ID condition that
limits the search scope. It is processed in the same way as a
multiple-keyword query is. To process a limited search query using
the multi-way join, site IDs or domain IDs are stored as text types
and the IR indexes are created for them (see Example 1).
Multiple-keyword or limited search queries take much longer
processing time than single-keyword queries do because they require
more disk accesses to find top-$k$ results from the postings having the common docID's
in multiple document lists.

As stated above, the query processing time varies
depending on the search condition type and the $k$ value of the
top-$k$ query. To simplify the queuing model, we transform every
query into one equivalent query type, i.e., the single-keyword
top-10 query. For example, let us assume that a single-keyword
top-10 query takes 2 ms while a multiple-keyword top-50 query
takes 4 ms in some system component. Then, we regard a multiple-keyword
top-50 query as two single-keyword top-10 queries for the component.

%
%
\subsubsection{Arrival Rate ($\lambda$)}
\label{subsubsec:Arrival_Rate}

Suppose the query arrival rate is  $\lambda$, and the requests are
uniformly distributed to the components of the same type. Then, the
arrival rate for a component is inversely proportional to the
number of the components of the type, i.e., the number of queues.
Table~\ref{tbl:Arrival_rates} shows the arrival rates for each
component where $nm$, $ncm$, $ns$,
and $nh$ represent the numbers of masters, of CPUs per master, of
slave nodes, and of network hubs, respectively. The queues for the
master CPUs and the master memory bus process one request per user
query while the queues for the network hubs process $ns$ requests
per user query because every slave processes the same query and
returns the results through the network hubs.

\vspace*{-0.3cm}
\renewcommand{\baselinestretch}{1.10}
\begin{table}[hbt]
\begin{center}
\caption{Arrival rates.} \label{tbl:Arrival_rates} \vspace*{0.1cm}
{\small  
\begin{tabular}[c]{|c|c|c|c|}
\hline
\multirow{2}{*}{Component}  & A master & A master   & A network \\
                & CPU      & memory bus &  hub \\
\hline
Arrival rate  & $ \displaystyle \frac{\lambda}{ncm \cdot nm}$ & $\displaystyle \frac{\lambda}{nm}$ & $\displaystyle \frac{ns}{nh}\lambda$ \\[5pt]
\hline
\end{tabular}
}
\end{center}
\end{table}
\renewcommand{\baselinestretch}{2.0}

%
%
\subsubsection{Weighted Arrival Rate ($\lambda$')}
\label{subsubsec:Weighted_Arrival_Rate}

\begin{figure*}
\centering
\epsfig{file=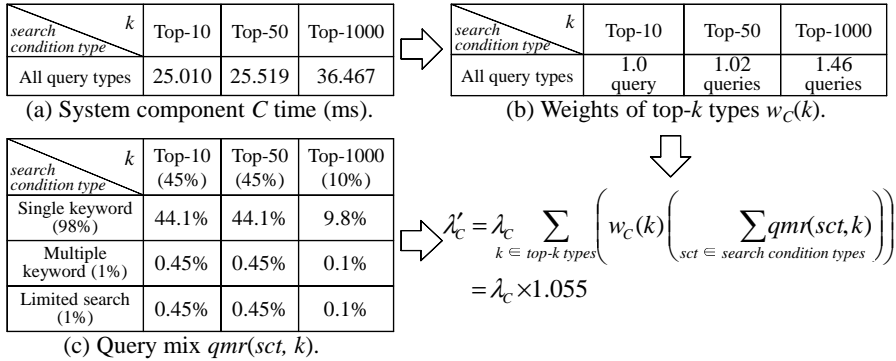, width=12cm}
\caption{An example of calculating a weighted arrival rate at the system component $C$.}
\label{fig:Weighted_arrival_rate_example}
\end{figure*}

In the query model, we transform a query of an arbitrary type to
an equivalent number of single-keyword top-10 queries. A
\textit{weighted arrival rate} of a component is the arrival rate
of the transformed queries. For each component, we measure the
processing time of each top-$k$ query and calculate the relative
weights of the processing time of top-50 and top-1000 queries
compared with that of a single-keyword top-10 query as the unit.
As an example, Fig.~\ref{fig:Weighted_arrival_rate_example}\,(a)
shows the average query processing time in the system component
$C$ of each top-$k$ query type, and
Fig.~\ref{fig:Weighted_arrival_rate_example}\,(b) shows the
weight $w_{C}(k)$ of each top-$k$ query type in $C$.
For the component $C$, the average processing time of a top-10
query is 25.01 ms, and its weight is considered to be 1.0. The
weights of other top-$k$ queries are calculated based on this
value. The weighted arrival rate at the component $C$, $\lambda_{C}$', is
obtained by calculating the sum of products of the weight and the
query mix ratio of each query type. For example, consider an example
of query mix ratio $qmr(sct, k)$ for top-$k$ queries
in Fig.~\ref{fig:Weighted_arrival_rate_example}\,(c), where $sct$
is the search condition type.
Then, the weight of the arrival rate at the system component $C$ for
this query mix is calculated as 1.055.

The weighted arrival rate at each system component is as in
Formulas~(\ref{eq:Weighted_arrival_rate_master-CPU})\,$\sim$\,(\ref{eq:Weighted_arrival_rate_network}).
Here, we separate master CPUs from the master memory bus since we
have multiple CPUs for each memory bus, and there is more
contention in the memory bus than in a CPU.
Since it is impossible to measure the weights of the master CPU and
the master memory bus separately, we measure the master's
total weight $w_{master}(k)$ for each top-$k$ type and assume the
same weight for both the master CPU and master memory bus.

{\small
\begin{eqnarray}
\lefteqn{ \lambda'_{master\mbox{-}CPU}(\lambda, nm, ncm) =  } \nonumber\\
& & \frac{\lambda}{ncm \cdot nm}
    \sum_{\substack{ k \in top\mbox{-}k\\types }}
    \biggl\{w_{master}(k)
    \sum_{\substack{ sct \in search\\condition\\types}}
    qmr(sct,k)
    \biggr\}
\label{eq:Weighted_arrival_rate_master-CPU}\\
\lefteqn{\lambda'_{master\mbox{-}memory\mbox{-}bus}(\lambda, nm) =  } \nonumber\\
& & \frac{\lambda}{nm}
    \sum_{\substack{k \in top\mbox{-}k\\types}}
    \biggl\{w_{master}(k)
    \sum_{\substack{sct \in search\\condition\\types}}
    qmr(sct,k)
    \biggr\}
\label{eq:Weighted_arrival_rate_master-memory}\\
\lefteqn{\lambda'_{network}(\lambda, ns, nh) =  } \nonumber\\
& & \frac{ns}{nh}\lambda
    \sum_{\substack{k \in top\mbox{-}k\\types}}
    \biggl\{w_{network}(k)
    \sum_{\substack{sct \in search\\condition\\types}}
    qmr(sct,k)
    \biggr\}
\label{eq:Weighted_arrival_rate_network}
\end{eqnarray}
}

%
%
\subsubsection{Parameters of the Performance Model}
\label{subsubsec:Parameters_of_the_Performance_Model}

To estimate a sojourn time at each component by using a queuing
model, we measure the service time of each component.
The service time at a component for each query type is obtained by
measuring and averaging the processing times of all the queries of
that type in a query set when executed alone. The master has
two types of components: CPUs and the memory bus. Since the master
CPU time and master memory access time cannot be measured
independently, we first measure the total time taken by a master
$ST_{master}(k, ns)$.
Then, we obtain the service time of a master CPU
$ST_{master\mbox{-}CPU}(k, ns)$ and that of the master memory bus
$ST_{master\mbox{-}memory\mbox{-}bus}(k, ns)$ by dividing the
total time by a given ratio, which will be discussed below. The
total service time of a master $ST_{master}(k, ns)$ is obtained by
Formula~(\ref{eq:Service time of master}).
In Formula~(\ref{eq:Service time of master}),
$T_{parent\mbox{-}proc}$ is the time spent by the parent process
in the master for checking syntax of the query.
$T_{child\mbox{-}proc}$ is the time spent by the child process for
distributing the query to the slaves and storing the results
received from the slaves into the result buffer.
$T_{master\mbox{-}RPC}(k)$ is the time taken by the
communication (RPC) module of the master; it varies according to $k$.
$T_{child\mbox{-}proc}$  and $T_{master\mbox{-}RPC}(k)$ are
multiplied by $ns$ since they are repeated as many times as the
number of slaves. $T_{merge}(k, ns)$ is the time for merging the
results from the slaves to get the final top-$k$ results.
$T_{context\mbox{-}switch}(k, ns)$ is the time for the master
processes to perform context switching. Formulas~(\ref{eq:Service
time of master-CPU}) and (\ref{eq:Service time of
master-memory}) show how to get the service times of a master CPU
and a master memory bus. The service time of a master measured,
$ST_{master}(k, ns)$, is divided according to the ratio of
$\alpha:(1-\alpha)$ to get the CPU time : the memory access time
($0 \leq \alpha \leq 1$). The value of $\alpha$ is chosen in such
a way that the estimated results fit the experimental results
actually measured using the five-node system.

{\small
\begin{eqnarray}
\lefteqn{ ST_{master}(k, ns) = } \nonumber\\
    & & T_{parent\mbox{-}proc} + (T_{child\mbox{-}proc} + T_{master\mbox{-}RPC}(k))*ns \nonumber\\
    & & + T_{merge}(k, ns) + T_{context\mbox{-}switch}(k, ns)
\label{eq:Service time of master} \\
\lefteqn{ ST_{master\mbox{-}CPU}(k, ns) = ST_{master}(k,
ns)\times\alpha}
\label{eq:Service time of master-CPU} \\
\lefteqn{ ST_{master\mbox{-}memory\mbox{-}bus}(k, ns) =
ST_{master}(k, ns)\times(1 - \alpha)} \label{eq:Service time of
master-memory}
\end{eqnarray}
}

In Formula~(\ref{eq:Service time of master}),
$T_{parent\mbox{-}proc}$, $T_{child\mbox{-}proc}$, and
$T_{master\mbox{-}RPC}(k)$ are measured using a small-sized
(five-node) system since these values are independent of the
number of slaves. On the other hand, since $T_{merge}(k, ns)$ and
$T_{context\mbox{-}switch}(k,ns)$ depend on the number of slaves,
we obtain them by using Formulas~(\ref{eq:Time_merge}) and
(\ref{eq:Time_master_context_switch}). To get the final top-$k$
results, the results from the slaves are merged using a loser
tree. In Formula~(\ref{eq:Time_merge}), $t_{comparison}$ is the time
spent by one comparison in the loser tree, and $\lceil \log_{2}ns
\rceil$, the height of the loser tree $- 1$, is the number of
comparisons for one result. $t_{base}$ is the time spent for
selecting a winner including the time to read streams and the time
to copy the result to the result buffer but excluding the time for
comparison. In Formula~(\ref{eq:Time_master_context_switch}),
$t_{per\mbox{-}context\mbox{-}switch}$ is the time spent by one
context switch in the master. $ncs_{base}(k)$ is the initial
number of context switches, and $ncs_{per\mbox{-}slave}(k)$ is the
additional number of context switches per slave.

{\small
\begin{eqnarray}
\lefteqn{ T_{merge}(k, ns) = k \times ( \lceil \log_{2}ns \rceil \times t_{comparison} + t_{base} ) } 
\label{eq:Time_merge}\\
\lefteqn{ T_{context\mbox{-}switch}(k, ns) = } \nonumber\\
    &  t_{per\mbox{-}context\mbox{-}switch} \times (ncs_{base}(k) + (ns \times ncs_{per\mbox{-}slave}(k)))
\label{eq:Time_master_context_switch}
\end{eqnarray}
}
\vspace*{0.30cm}

\begin{figure}[hbt]
  \centerline{\epsfig{file=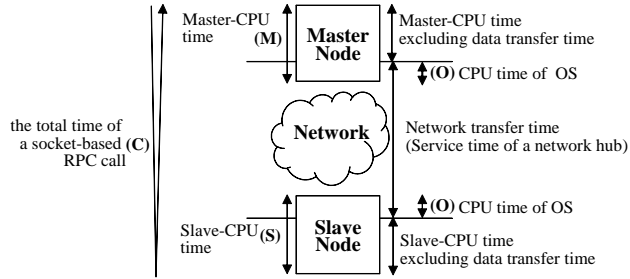, width=8.5cm}}
  \caption{Measurement of a service time at a network hub.}
  \label{fig:Service_time_of_network}
\end{figure}

$ST_{network}(k)$, the service time of a network hub, is obtained
by measuring the time taken by a network hub to transfer top-$k$
results of a slave. Fig.~\ref{fig:Service_time_of_network} shows
how to measure the service time of a network hub. For each top-$k$
result, we measure the total time C of an RPC call. We then
subtract M and S from C, where M is the CPU time taken by a master
node, and S is the CPU time taken by a slave node. The CPU times
are measured by using the time utility of LINUX. The CPU times M
and S overlap the time O, which is the CPU time of the operating
system to transfer data. However, the time is spent concurrently
with the network transfer time, so we measure the time O and add
it back. To measure the time O, we make a dummy RPC function by
removing data transfer from the original RPC function, and measure
the CPU time of the master node. The measured value corresponds to
M$-$O, so we can obtain O from M and M$-$O. We assume that the CPU
time of the operating system for network transfer at a slave is
the same as that at the master. Therefore, we measure the service time
of a network hub as (C$-$M$-$S)$+$2$\cdot$O.

%
%
\subsubsection{Average Queue Length and the Estimated Sojourn Time }
\label{subsubsec:Average_Queue_Length_and_the_Estimated_Sojourn_Time}

The average queue length of a component is the sum of the number
of customers waiting in the queue and the number of customers
being serviced. The sojourn time, i.e., the response time of a
queue, is determined by the length of the queue. In this paper,
every queue used in the performance model is an M/D/1 queue with a
Poisson arrival process, a constant service time, and a single
server. The average length of an M/D/1 queue is obtained by
Formula~(\ref{eq:Average_queue_length_MD1}). The average queue
length for each component is obtained by
Formulas~(\ref{eq:Average_queue_length_master-CPU})\,$\sim$\,(\ref{eq:Average_queue_length_network}).
To obtain the average queue length for the component $C$, we
substitute $\lambda$ in
Formula~(\ref{eq:Average_queue_length_MD1}) with the weighted
arrival rate, and $ST$ with the service time of top-10 query type
for the component because the queries are normalized in terms of
the single keyword top-10 query. We use a fixed value of $ST$,
which is the average service time for the queue.

{\small \setlength\arraycolsep{2pt}
\begin{eqnarray}
\lefteqn{ L(\lambda, ST) = \frac{\lambda^2 E[ST^2]}{2(1-\lambda
E[ST])} + \lambda E[ST] }
\label{eq:Average_queue_length_MD1}\\
\lefteqn{ L_{master\mbox{-}CPU}(\lambda, nm, ncm, ns) = } \nonumber\\
& & L(\lambda'_{master\mbox{-}CPU}(\lambda, nm, ncm), ST_{master\mbox{-}CPU}(k=10, ns)) \nonumber\\
\label{eq:Average_queue_length_master-CPU}\\
\lefteqn{ L_{master\mbox{-}memory\mbox{-}bus}(\lambda, nm, ns) = } \nonumber\\
& & L(\lambda'_{master\mbox{-}memory\mbox{-}bus}(\lambda, nm), ST_{master\mbox{-}memory\mbox{-}bus}(k=10, ns)) \nonumber\\
\label{eq:Average_queue_length_master-memory}\\
\lefteqn{ L_{network}(\lambda, ns, nh) = } \nonumber\\
& & L(\lambda'_{network}(\lambda, ns, nh), ST_{network}(k=10))
\label{eq:Average_queue_length_network}
\end{eqnarray}
}

Meanwhile, the sojourn time $X$ of a customer in a queue is the
total time the customer spends inside the queue for waiting and
for being served. The average sojourn time is obtained by
Formula~(\ref{eq:Expected_sorjourn_time}), where $L$ is the average queue
length and $\lambda$ the arrival rate. Thus, the average
sojourn time of single-keyword top-10 queries for each component
is obtained by dividing the average length of the corresponding
queue by the weighted arrival rate. The average sojourn times of
queries having \textit{other} top-$k$ values are obtained by
multiplying the weight for the top-$k$ of the component. For a
network hub, $ns/nh$ is additionally multiplied because $ns/nh$
slaves are connected to one network hub.

{\small
\begin{eqnarray}
\lefteqn{ E[X]=\frac{L}{\lambda} }
\label{eq:Expected_sorjourn_time}\\
\lefteqn{ E[X_{master\mbox{-}CPU}]= } \nonumber\\
& & \frac{L_{master\mbox{-}CPU}(\lambda, nm, ncm,
ns)}{\lambda'_{master\mbox{-}CPU}(\lambda, nm, ncm)}\times
w_{master}(k)
\label{eq:Expected_sorjourn_time_of_master-CPU}\\
\lefteqn{ E[X_{master\mbox{-}memory\mbox{-}bus}]= } \nonumber\\
& & \frac{L_{master\mbox{-}memory\mbox{-}bus}(\lambda, nm,
ns)}{\lambda'_{master\mbox{-}memory\mbox{-}bus}(\lambda,
nm)}\times w_{master}(k)
\label{eq:Expected_sorjourn_time_of_master-memory}\\
\lefteqn{ E[X_{network}]= } \nonumber\\
& & \frac{ns}{nh}\times
    \frac{L_{network}(\lambda, ns, nh)}{\lambda'_{network}(\lambda, ns, nh)}\times w_{network}(k)
\label{eq:Expected_sorjourn_time_of_network}
\end{eqnarray}
}
\vspace*{-0.50cm}

%
%
\subsection{Estimation of the Expected Slave Max Time}
\label{subsec:The_Estimation_Method_of_the_Expected_Slave_Max_Time}

In ODYS, given a user query, all of the slaves process the query
in parallel at semi-cold start. When a query is executed at
semi-cold start, different slaves have much different processing
times because the disk access time has a lot of variation.
We note that, the total processing time is bounded by the maximum slave
sojourn time (briefly, \textit{slave max time}) since we must
receive the results from all the slaves to answer the query.
Unfortunately, it is known to be difficult to get analytically the
expected value of the maximum sojourn time for multiple
queues\,\cite{Kem09}. Thus, we propose a method for estimating the
slave max time based on measurement.

\begin{figure}[hbt]
  \centerline{\epsfig{file=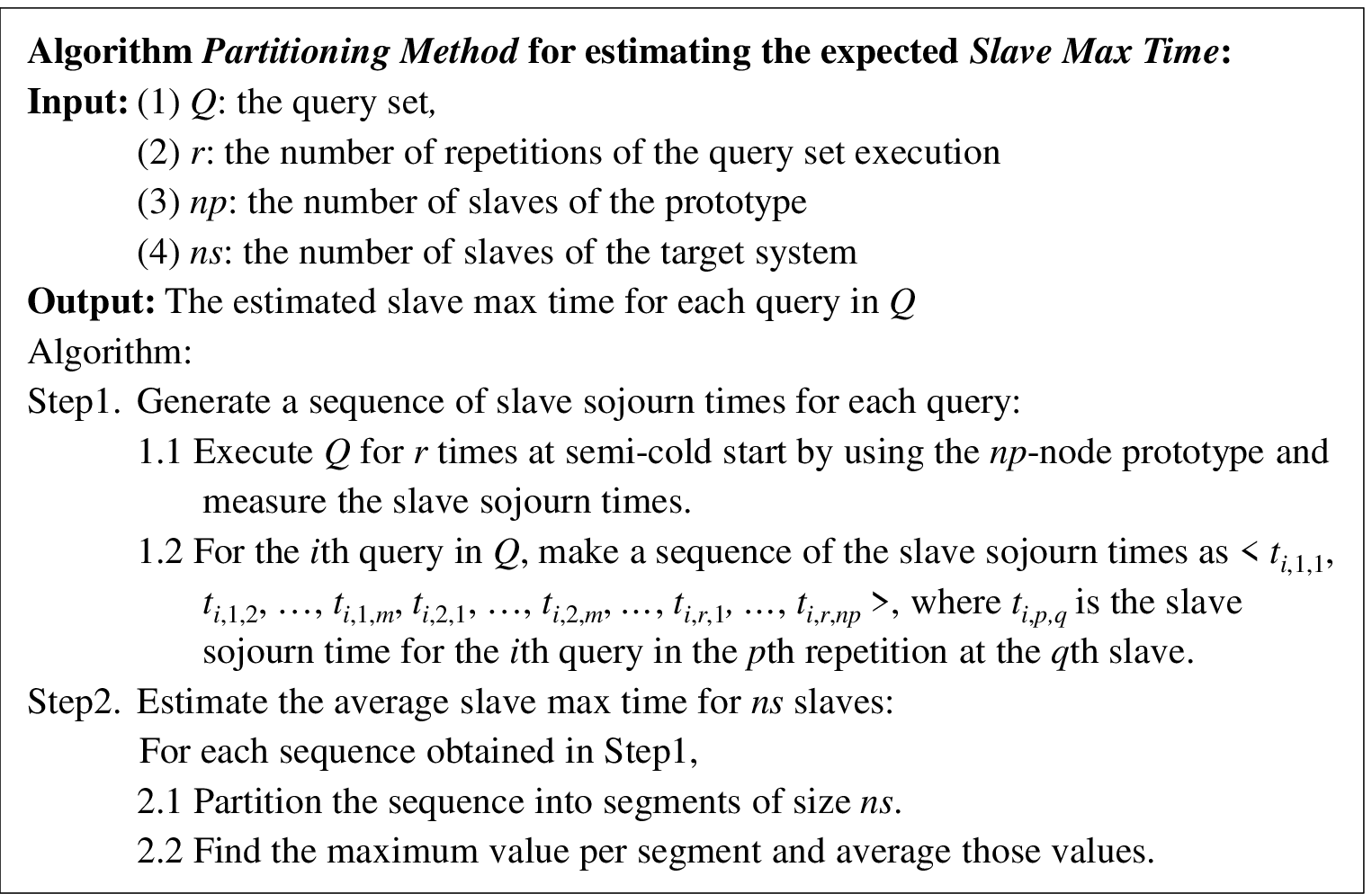, width=10cm}}
  \caption{The algorithm for estimating the expected slave max time by using the partitioning method.}
  \label{fig:Alg_of_partitioning_method}
\end{figure}

Fig.~\ref{fig:Alg_of_partitioning_method} shows the algorithm of
the proposed estimation method for the slave max time. We call
this algorithm the \textit{partitioning-based estimation method}
(simply, the {\it partitioning method}). The partitioning method estimates the slave max time of a
large-sized (e.g., 300-node) target system by running a
small-sized (e.g., 5-node) test system multiple times repeatedly.
Suppose $r$ is the number of repetitions, $np$ the number of slaves
in the test system, and $ns$ the number of slaves in the
target system. In Step 1, the algorithm generates a sequence
of $np\times r$ slave sojourn times for each query by running the test
system $r$ times. In Step 2, the algorithm partitions the sequence
into segments of size $ns$, find the maximum value per segment, and
average these values. Since the partitioning method provides the maximum
value by measurement, the result must be very close to the actual measurement
of the target system.

%
%
\subsection{Estimation of the Average Total Query Response Time}
\label{subsec:The_Estimation_Method_of_the_Average_Response_Time}
\vspace*{-0.40cm}

\begin{figure}[hbt]
  \centerline{\epsfig{file=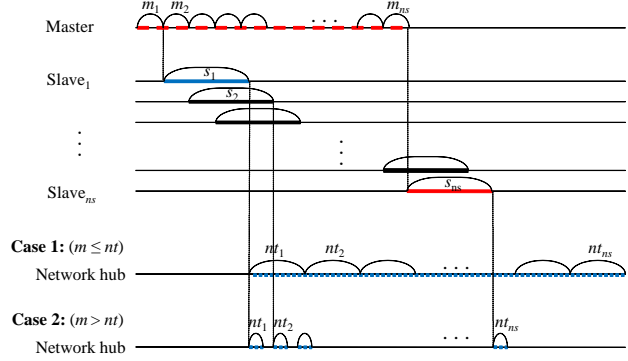, width=9cm}}
  \caption{Overview of estimating the average total query response time.}
  \label{fig:Query_processing_time_overview}
\end{figure}

Fig.~\ref{fig:Query_processing_time_overview} shows an overview
of the estimation method for computing the average total query
response time. In Fig.~\ref{fig:Query_processing_time_overview},
$m_{i}$ is the master processing time for a slave (i.e., for
sending the query and receiving the results from a slave),
$s_{i}$, the query processing time of the $i$th slave, and
$nt_{i}$, the network transfer time of the $i$th slave's result.
Suppose that for all $i$, $m_{i}$, $s_{i}$, and $nt_{i}$ have the same value
$m$, $s$, and $nt$, respectively.
Then, there are two cases for estimating the response time of a
query. If $m$ is less than or equal to $nt$, the total query
response time is obtained as $ (m_{1} + s_{1} + ns\times nt) = (m
+ s + ns\times nt) \approx ( ns \times nt + s)$. Or, if $m$ is
greater than $nt$, the total query response time is obtained as
$ (ns \times m + s_{ns} + nt_{ns}) = (ns \times m + s + nt)
\approx (ns \times m + s) $. (One $m$ or one $nt$ is negligible.)
In other words, the total query response time is obtained by
adding the bigger of the network hub's total sojourn time
$(ns\times nt)$ and the master's total sojourn time $(ns \times
m)$ to the slave sojourn time $(s)$. Here, for $s$,
we use the slave max time as discussed in
Section~\ref{subsec:The_Estimation_Method_of_the_Expected_Slave_Max_Time}.
Formula~(\ref{eq:Query_processing_time}) shows how to get the
estimated value of the average total query response time.

\begin{figure*}[hbt]
{\small
\begin{eqnarray}
\lefteqn{ t_{parallel\mbox{-}n\mbox{-}node}(sct, k, \lambda, nm,
ncm, ns, nh) }
\nonumber\\
    & = & max \biggl( \biggl(E[X_{master\mbox{-}CPU}] + E[X_{master\mbox{-}memory\mbox{-}bus}]\biggr), E[X_{network}] \biggr)
    + t_{slave\mbox{-}max\mbox{-}time}(sct, k, \lambda, ns)
\nonumber\\
    & = & max \biggl(
                \biggl(
                    \frac{L_{master\mbox{-}CPU}(\lambda, nm, ncm, ns)}{\lambda'_{master\mbox{-}CPU}(\lambda, nm, ncm)}\times w_{master}(k)
                    + \frac{L_{master\mbox{-}memory\mbox{-}bus}(\lambda, nm, ns)}{\lambda'_{master\mbox{-}memory\mbox{-}bus}(\lambda, nm)}\times w_{master}(k)
                \biggr),
\nonumber\\
                & & \frac{ns}{nh}\times \frac{L_{network}(\lambda, ns, nh)}{\lambda'_{network}(\lambda, ns, nh)}\times w_{network}(k)
            \biggr)
    + t_{slave\mbox{-}max\mbox{-}time}(sct, k, \lambda, ns)
    \label{eq:Query_processing_time}
\end{eqnarray}
}
\end{figure*}

\section{Performance Evaluation}
\label{sec:Performance_Evaluation}

In this section, we validate the performance model and project the
performance of a real-world-scale ODYS using this model.

%
%
\subsection{Experimental Data and Environment}
\label{subsec:Experimental_Data_and_Environment}

We have built a five-node parallel search engine according to the
architecture in Fig.~\ref{fig:Architecture}. Here, we have used
one master, one 1-Gbps hub, and five slaves each running 100
Odysseus processes with a shared buffer of 12 Mbytes. We have used
one machine for the master, a Linux machine with a quad-core 3.06
GHz CPU and 6 Gbytes of main memory. We have used five machines
for the slaves. Four of them are Linux machines with two dual-core
3 GHz CPU, 4Gbytes of main memory, and a RAID 5 disk array. Each
disk array has 13 disks (disk transfer rate: average 59.5
Mbytes/s) with a total of 0.9\,$\sim$\,3.9 Tbytes disk space, a
cache of 0.5\,$\sim$\,1 Gbytes, and 200 Mbytes/s bandwidth. The
remaining slave machine is a Linux machine with a quad-core 2.5
GHz CPU, 4 Gbytes of main memory, and a RAID 5 disk array. The
disk array has 13 disks (disk transfer rate: average 83.3
Mbytes/s) with a total of 13 Tbytes, a cache of 512 Mbytes, and
512 Mbytes/s bandwidth. The master and the slaves are connected by
a 1-Gbps network hub.

%
%
We perform experiments using 114 million Web pages crawled. For
each Web page, if its size is larger than 16 Kbytes,
we extract and store important part of a fixed size
(front 8 KBytes + rear 8 KBytes) to uniformly control the
experiments. We evenly partition the set of Web pages into five
segments and allocate each segment to a slave. Thus, each slave
stores and indexes 22.8 million Web pages\footnote{One ODYS slave
is capable of indexing 100 million Web pages. Here, we used 22.8
million Web pages/slave since we had only 114 million Web pages
crawled. However, this does not affect the query performance since
the postings are sorted according to the PageRank order, and only
up to 1000 postings are retrieved for a single-keyword top-$k$
query. For a multiple-keyword or limited search query, a larger
number of postings are accessed, but a sufficient number of
postings is available from 22.8 million Web pages.}. We perform
experiments using two query sets. One consists of only
single-keyword top-10 queries, and the other consists of a mixed
type of queries with different $k$ (in top-$k$) values. We call
the query sets SINGLE-10-ONLY and QUERY-MIX, respectively. For
QUERY-MIX, we use the query mix ratio in
Fig.~\ref{fig:Weighted_arrival_rate_example}\,(c). Each query
set includes 10,000 queries in which the keywords, site IDs, and
domain IDs are all unique. Queries are generated at a Poisson
arrival rate and issued by a separate machine.

%
%
We perform the following experiments. First, we measure the
estimation error of the performance model by comparing its
projected output with the results measured using the five-node
system. We use the \textit{estimation error} as defined in
Formula~(\ref{eq:Error_rate_of_PEM}). Second, we estimate the
slave max time using the partitioning method.
Last, we project the query processing performance of ODYS for
real-world-scale data and query loads by using the performance
model. The value of $\alpha$ used in the performance model is
0.25. All the experimental results are measured at semi-cold
start. To measure the semi-cold start times, we first empty the
DBMS buffers and disk array caches of all the slaves, and then,
run 10,000 queries that are completely independent of the
experimental queries (i.e., no overlapping keywords, site IDs, or
domain IDs with the experimental queries) to load the internal
nodes of the IR index into the DBMS buffer.

\renewcommand{\baselinestretch}{1.0}
{\small
\begin{eqnarray}
\begin{array}{c} { \emph{estimation}} \\ { \emph{error}} \end{array} =
    \frac{ \left|
        \begin{array}{c} { \emph{estimated average}} \\ { \emph{time of a query}} \end{array}
        -
        \begin{array}{c} { \emph{measured average}} \\ { \emph{time of a query}} \end{array}
    \right|}
    {
        \begin{array}{c} { \emph{measured average}} \\ { \emph{time of a query}} \end{array}
    }
\label{eq:Error_rate_of_PEM}
\end{eqnarray}
}
\renewcommand{\baselinestretch}{2.0}

%
%
\subsection{Experimental Results}
\label{subsec:Experimental_Results}

%
%
\subsubsection{Measurement of Parameters}
\label{subsubsec:Measurement_of_Parameters}

We measure the parameters of the queuing model using the five-node
system. The values are measured as described in
Section~\ref{subsubsec:Parameters_of_the_Performance_Model}.
Table~\ref{tbl:Measurement_of_parameters} shows the parameters measured.

\renewcommand{\baselinestretch}{1.10}
\begin{table}[hbt]
\begin{center}
\caption{Parameters of the queuing model measured.}
\vspace*{0.1cm} \label{tbl:Measurement_of_parameters}
{\small  
\begin{tabular}[c]{|c|c|} \hline
Parameters & Values  \\
\hline \hline
$T_{parent\mbox{-}proc}$  & 1.516 ms  \\
\hline
$T_{child\mbox{-}proc}$  & 0.0181 ms  \\
\hline
\multirow{3}{*}{ $T_{master\mbox{-}RPC}(k)$ } & 0.01 ms, $k$ = 10\\
    & 0.011 ms, $k$ = 50\\
    & 0.031 ms, $k$ = 1000\\
\hline
$t_{comparison}$ & 0.191 $\mu$s\\
\hline
$t_{base}$ & 0.28 $\mu$s \\
\hline
$t_{per\mbox{-}context\mbox{-}switch}$ & 15.995 $\mu$s \\
\hline
\multirow{2}{*}{$ncs_{base}(k)$}  & 80.869, $k$ = 10, 50\\
                            & 139.903, $k$ = 1000\\
\hline
\multirow{2}{*}{$ncs_{per\mbox{-}slave}(k)$ } & 1.991, $k$ = 10, 50\\
                            & 3.444, $k$ = 1000\\
\hline
\multirow{3}{*}{ $ST_{network}(k)$ } & 0.129 ms, $k$ = 10\\
    & 0.222 ms, $k$ = 50\\
    & 0.318 ms, $k$ = 1000\\
\hline
\end{tabular}
} 
\end{center}
\vspace*{-0.7cm} 
\end{table}
\renewcommand{\baselinestretch}{2.0}

%
%
\subsubsection{Accuracy of the Performance Model}
\label{subsubsec:Accuracy_of_the_Performance_Model}

\begin{figure}[hbt]
\centerline{\epsfig{file=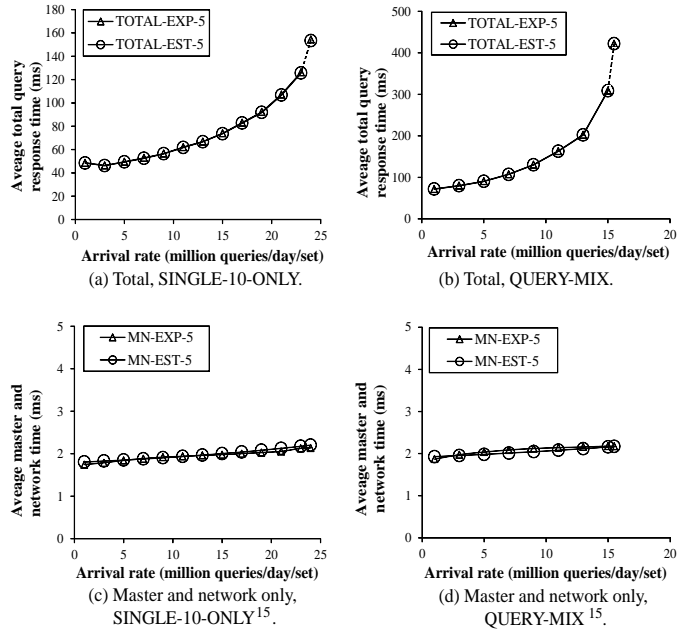,
width=9cm, height=8.5cm}} \caption{The estimated and experimental
results of the five-node system ($ns$=5). TOTAL-EXP-5 and MN-EXP-5
represent experimental results; TOTAL-EST-5 and MN-EST-5 estimated
results.} \label{fig:Exp_real_world_simulation}
\end{figure}

We vary the query loads from 1 to 24 million queries/day for the
two query sets: SINGLE-10-ONLY and QUERY-MIX.
Fig.~\ref{fig:Exp_real_world_simulation} shows the average
total query response time and the average processing time of the
master and network (i.e., the part modeled by the queuing model)
for each query load. In Fig.~\ref{fig:Exp_real_world_simulation},
TOTAL-EXP-5 represents the experimental results of the total query
response time measured using the five-node system; TOTAL-EST-5
represents the estimated results of the total query response time
from the performance model. MN-EXP-5 represents the experimental
results of the master and network time; MN-EST-5 represents the
estimated results. The results show that the maximum estimation error
of the total query response time is 0.59\% for SINGLE-10-ONLY, and
0.50\% for QUERY-MIX. The maximum estimation error of the queuing
model including the master and network only is 3.41\% for SINGLE-10-ONLY
and 3.62\% for QUERY-MIX\footnote{For sensitivity analysis, we have tested a different query mix having
20\% of top-1000 queries obtaining a similar result where the
maximum estimation error was 5.33\%. We have not conducted sensitivity
analysis on search condition types since the master and network time
does not depend on search condition types, but only on the top-$k$ value.}.
For the same query load, the result of SINGLE-10-ONLY and QUERY-MIX shows a
big difference. The reason is that the response time of the multiple-keyword and
limited search queries are much longer than those of the
single-keyword queries as discussed in Section~\ref{subsubsec:Query_Model}\footnote{It can
be optimized by constructing separate ODYS sets dedicated to limited search queries.
For the IR indexes of these ODYS sets, we can order the postings of each
posting list by the domain ID, and then, by ranking the sets of
the postings that have the same domain ID's independently of one
another, significantly reducing the search over the posting list.
However, these additional optimizations are beyond the
scope of this paper and will be left as further study.}.
In Fig.~\ref{fig:Exp_real_world_simulation}, dotted lines represent
the regions where measurements become unstable since the number of
input queries becomes close to the maximum possible throughput.
The result in Fig.~\ref{fig:Exp_real_world_simulation}\,(a) shows that the 5-node ODYS
can stably process 266 queries/sec (23 million queries/day) with an average response time of
126 ms on a 780 GBytes dataset.

\renewcommand{\baselinestretch}{1.0}
\addtocounter{footnote}{1} \footnotetext{For
Figures~\ref{fig:Exp_real_world_simulation}\,(c) and (d), the master and network
time is measured by subtracting the slave max time from the total
query response time.}
\renewcommand{\baselinestretch}{2.0}

%
%
\subsubsection{Estimation of the Slave Max Time}
\label{subsubsec:Estimation_Results_of_the_Slave_Max_Time}

\begin{figure}[hbt]
  \centerline{\epsfig{file=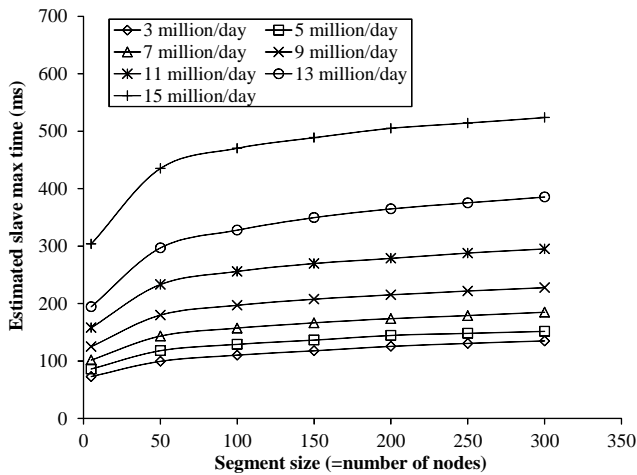, width=8.5cm}}
  \caption{The estimated slave max time as the segment size is varied (QUERY-MIX, $r$=5, $ns$=5).}
  \label{fig:Exp_slave_max_time}
\end{figure}

We analyze the effect of the number of slaves on the performance
of ODYS. We estimate the slave max times for QUERY-MIX while
increasing the segment size of the partitioning method for each
query load. We measure 300 slave sojourn times for each query by
running the query set 60 times using the five-node system.
Fig.~\ref{fig:Exp_slave_max_time} shows the expected slave max
time for each segment size. The results show that the expected
slave max time increases up to 1.5\,$\sim$\,2 times of the minimum
value as the segment size increases, i.e., as the number of slaves
increases. Interestingly, the slave max time gradually converges
to a value less than twice the minimum instead of increasing
indefinitely. Detailed analysis of this phenomenon is beyond the scope of this
paper. We leave it as a further study.

%
%
\subsubsection{Performance Projection of a Real-World-Scale (300-Node) ODYS}
\label{subsubsec:Performance_Projection_of_a_Real-World-Scale_ODYS}

\begin{figure}[hbt]
  \centerline{\epsfig{file=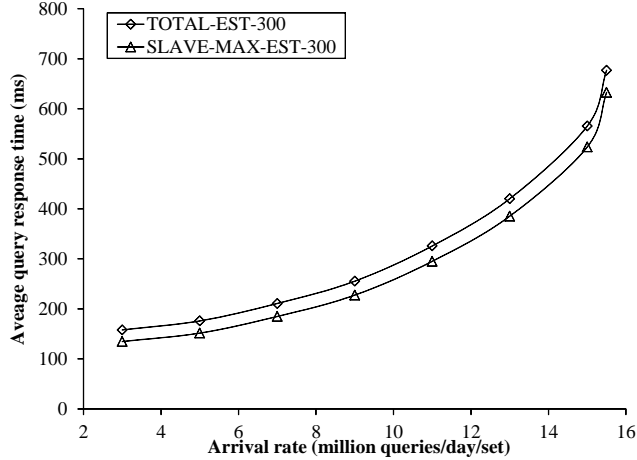, width=8.5cm}}
  \caption{The projected average response time of ODYS for real-world-scale service ($ns$=300).
  TOTAL-EST-300 is the average estimated total query response time; SLAVE-MAX-EST-300 the average estimated slave max time.}
  \label{fig:Exp_real_world_scale_estimation}
\end{figure}

By using the performance model, we estimate the average response
time of ODYS for 30 billion\footnote{Google (as well as other
commercial search engines such as MS Bing and Yahoo!) indexes
approximately 25 billion Web pages. This is roughly indicated by
the result of querying with the frequently occurring keywords such
as ``a,'' ``the,'' or ``www.''} Web pages, which is considered
real-world-scale data\,\cite{Gul05,Kun}.
Fig.~\ref{fig:Exp_real_world_scale_estimation} shows the
estimated performance of a 300-node system for the query set
QUERY-MIX. In the estimation, one ODYS set consists of 4 masters,
300 slaves (a slave is capable of indexing 100 million Web pages),
and 11 Gbit network hubs. Each master has a Quad-Core 3.06 GHz CPU,
each slave has one 3 GHz CPU, 4 Gbytes of main memory,
and $13 \times 300$ Gbytes SATA hard disks.
Here, we select the number of masters (4) and network hubs (11) to
make the queue lengths of master memory and network hubs similar
to each other to avoid bottlenecks.
In Fig.~\ref{fig:Exp_real_world_scale_estimation}, TOTAL-EST-300
represents the estimated total response time of queries, and
SLAVE-MAX-EST-300 represents the slave max time estimated using
the partitioning method. SLAVE-MAX-EST-300 is identical to the
results of the segment size 300 in
Fig.~\ref{fig:Exp_slave_max_time}.

As we observed in Fig.~\ref{fig:Exp_real_world_scale_estimation},
if an ODYS set were to take a higher query load, the total number
of nodes required to handle the load could be reduced, but the
performance would be degraded. Therefore, the trade-off between
the number of nodes and the performance exists in providing
reasonable performance with a minimal number of nodes. For
example, suppose we run 7 million queries/day (81 queries/sec) per ODYS set,
then ODYS can handle Google-scale service using 143 sets of 304 nodes
(4 masters and 300 slaves), a total of 43,472 nodes, for 1 billion
queries/day. In this case, the average query response time is only
211 ms. In contrast, if we ran 3.5 million queries/day (40.5 queries/sec),
ODYS would need 286 sets with a total of 86,944 nodes with an average query
response time of 162 ms. The commercial search engine implemented
on a distributed file system uses hundreds of thousand of
nodes\,\cite{Dea09b}, and the average response time of such a system
is known to be 200-250 ms\,\cite{Dea09,Goo}. The experiments show
that our approach is capable of providing commercial-level service
with a much smaller number of nodes than a DFS-based approach.

\section{Conclusions}
\label{sec:Conclusions}

In this paper, we have shown that a massively parallel search
engine capable of processing real-world-scale data and
query loads can be implemented using a DB-IR tightly integrated
parallel DBMS. We have presented the architecture of a massively
parallel search engine, ODYS, using such a DBMS. The DBMS used in
the slaves of ODYS is Odysseus, which is a highly scalable ORDBMS
that is tightly integrated with IR features\,\cite{Wha05}.
Odysseus is capable of indexing 100 million Web pages, and thus,
ODYS is able to handle a large volume of data even with a small
number of machines. The tightly integrated DB-IR functionalities
enable ODYS to have commercial-level performance for keyword
queries. Furthermore, ODYS provides rich functionality such as
SQL, schemas, and indexes for easy (and less error-prone) development
and maintenance of applications\,\cite{Wha11}.

We have also proposed a performance model that
can project the performance of the proposed architecture.
The model is a hybrid one that employs both an analytic approach
based on the queuing model and an experimental approach of using
a small-sized test system to project the performance of a large target system.
We have validated this model through comparison of the result projected
by the model with the results measured using the five-node system.
Our estimation of the total response time is quite accurate
since the bulk of the total response time is spent at the slave, and we derive the slave max time
by measurement with accuracy. The comparison between the estimated and experimental results
indicates that the estimation error of the total query
response time of the five-node system is less than 0.59\%.
Such a modeling method is helpful in realistically estimating the
performance of a system by using limited resources---without
actually building a large-scale system.

\newpage
Finally, we have estimated the performance of ODYS for
real-world-scale data and query loads. According to the
performance model, with a relatively small number of (i.e.,
43,472) nodes, ODYS is capable of handling 1 billion queries/day
for 30 billion Web pages at an average query response time of 211
ms. With twice as many nodes (i.e., half the query load per node),
it can provide an average query response time of 162 ms. These
results clearly demonstrate superior performance and scalability
of the proposed architecture compared with commercial search
engines using hundreds of thousands of nodes at an average query
response time of 200-250 ms\,\cite{Dea09,Goo}. These results are
even more marked since the performance of ODYS was evaluated at semi-cold start
with only 12 Mbytes of buffer while commercial search engines
heavily rely on in-memory processing by using massive-scale
main memory to store all the indexes in the buffer\,\cite{Dea09}.


\end{document}